\documentclass{mem}
\usepackage{natbib}\usepackage{txfonts}\usepackage{balance}
\usepackage{graphicx}
\usepackage[a4paper]{hyperref}

\newcommand{\xmm}{{\em XMM-Newton}}

\newcommand{\kms}{~km~s$^{-1}$}

\newcommand {\be}{\begin {equation}}
\newcommand {\ee}{\end {equation}}

\idline{75}{282}
\begin{document}
\def\teff{$T\rm_{eff }$}
\def\kms{$\mathrm {km s}^{-1}$}

\title{
Gas Turbulent Motions in Galaxy Clusters
}

   \subtitle{}

\author{
I. \,Zhuravleva\inst{1} 
          }

  \offprints{I. Zhuravleva}

\institute{
Max Planck Institute for Astrophysics, 
Karl-Schwarzschild str. 1, 85741, Garching, Germany
\email{izhur@mpa-garching.mpg.de}
}

\authorrunning{Zhuravleva}

\titlerunning{Turbulence in Galaxy Clusters}

\abstract{
We discuss various possibilities to constrain ICM turbulence in galaxy 
clusters using bright X-ray lines. Numerical simulations are used to 
find the most appropriate description of the 3D velocity field power 
spectrum (PDS) and to calibrate the relation of observables to this PDS.
The impact of the velocity field on the surface brightness distribution and 
on the spectral shape of strong X-ray lines, modified by the resonant 
scattering (RS), is evaluated via radiative transfer calculations.  We
investigate the sensitivity of RS not only to
amplitudes of motions, but also to anisotropy and spatial scales.  We
in particular show that the amplitude of radial 
motions is most important for RS, while tangential 
motions only weakly affect the scattering.

}
\maketitle{}

\section{Introduction}

The dynamical state of the hot gas in galaxy clusters
is still little known. It is believed that as clusters merge or as
matter accretes along filaments, turbulence should set up in the ICM,
energy of which should cascade down to the small scales and
dissipate. However resolution of numerical simulations is not yet sufficient to 
fully resolve this process.

Current observations give us only upper limits on turbulence in
clusters by means of direct measurements of line width \citep{San10}
or by resonant scattering (RS) measurements \citep{Chu04, Wer09}. 
In the nearest future {\em Astro-H}
observatory with its high energy resolution will allow us to measure
shifts and width of lines with high accuracy.

Here we discuss a way to relate observables, such as line-of-sight velocity
and dispersion, with full 3D velocity PDS. First we
consider PDS from simulated galaxy clusters, then we
illustrate how to implement calibration and at the end we consider
RS as one of the ways to measure and study velocities of gas motions.

\section{3D Velocity Power Spectrum and Observables}

\begin{figure}
\begin{minipage}{0.48\textwidth}
 \includegraphics[width=0.85\textwidth]{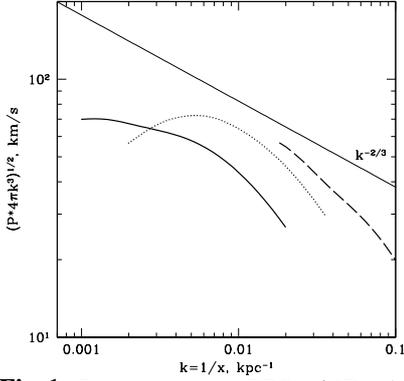}
\vspace{-5mm}
\caption{
\footnotesize
Power spectrum (PDS) of 3D velocity field in galaxy cluster from SPH simulations. PDS are  calculated in cubes of 1 Mpc (solid curves),
 0.5 Mpc (dotted curves) and 55 kpc (dashed curves) on a side through
  the variance method described in \cite{Are10}. Kolmogorov PDS is
  shown with the thin solid line.
}
\label{fig:3dpds} 
\end{minipage}   
\end{figure}

Hydrodynamical simulations provide us information about full 3D
velocity field in galaxy clusters. Using results of SPH
simulations \citep{Dol08} we calculated PDS (Fig. \ref{fig:3dpds})
for cluster g676, which has the highest resolution, 
  through the method described by
 \cite{Are10} which avoids a problem of non-periodic boundaries
 in data boxes. One can see (i) a significant deviation of the PDS
 shape from the canonical Kolmogorov PDS, (ii) that both shape and amplitude of PDS depend on
 the size of the cube used for calculation.
 To some extent this behavior could be attributed to insufficient (and 
distance dependent) resolution of SPH simulation and to the artificial 
viscosity.

\begin{figure}
\begin{minipage}{0.48\textwidth}
 \includegraphics[width=0.85\textwidth]{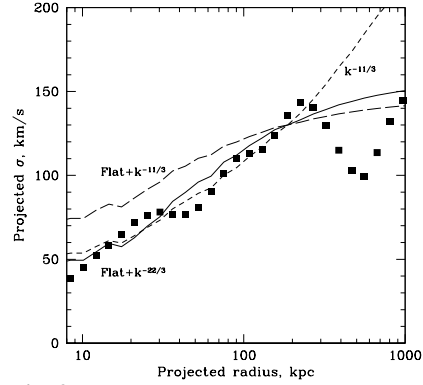}
\vspace{-5mm}
\caption{
\footnotesize
Profile of projected velocity dispersion for simulated galaxy cluster
(dots). Long-dashed, short-dashed and solid curves show velosity
dispersion calculated from eq. (\ref{eq:sig_pds}) assuming three
different PDS of velocity field. Notice that measured
projected velocity dispersion (emission measure weighted mean) can be
considered as a structure function since the interval with size $\sim
R$ mostly contribute to the dispersion measured at the projected distance $R$.}
\label{fig:sig_pds} 
\end{minipage}   
\end{figure}

Since gas motions are predominantly subsonic, the centroid shift and the 
width of lines, measured with X-ray observatories, contain most 
essential information on the ICM velocity field \citep{Ino03}. I.e. we have information about
emission measure weighted mean velocity and dispersion along the line
of sight. Relations between these observables and the 3D velocity PDS 
are the following:
\be
P_{2D}(k)=\int P_{3D}(\sqrt{k^2+k_z^2})W^2(k_z)dk_z, 
\label{eq:vel_pds}
\ee
\be
\sigma^2=\int P_{3D}(1-W^2(k_z))dk_zdk_ydk_x,
\label{eq:sig_pds}
\ee
where $P_{2D}(k)$ is a PDS of observed mean velocity,
$\sigma$ is observed velocity dispersion, $P_{3D}$ is a 3D PDS and 
$W$ is a Fourier transform of $n_e^2(z)$. In
Fig. \ref{fig:sig_pds} we illustrate the relation (\ref{eq:sig_pds})
for simulated galaxy cluster, PDS of which is shown in
Fig. \ref{fig:3dpds}. One can see that velocity dispersion from
``flat$+k^{-22/3}$" PDS model (flat at low $k$ and twice steeper than
Kolmogorov PDS at higher $k$) fits the observed
velocity dispersion very good, which is in agreement with the PDS shown in Fig. \ref{fig:3dpds}.

\section{Resonant Scattering as a Way to Measure Gas Motions}

\begin{figure*}
\begin{minipage}{0.38\textwidth}
  \includegraphics[width=0.8\textwidth,angle=270.]{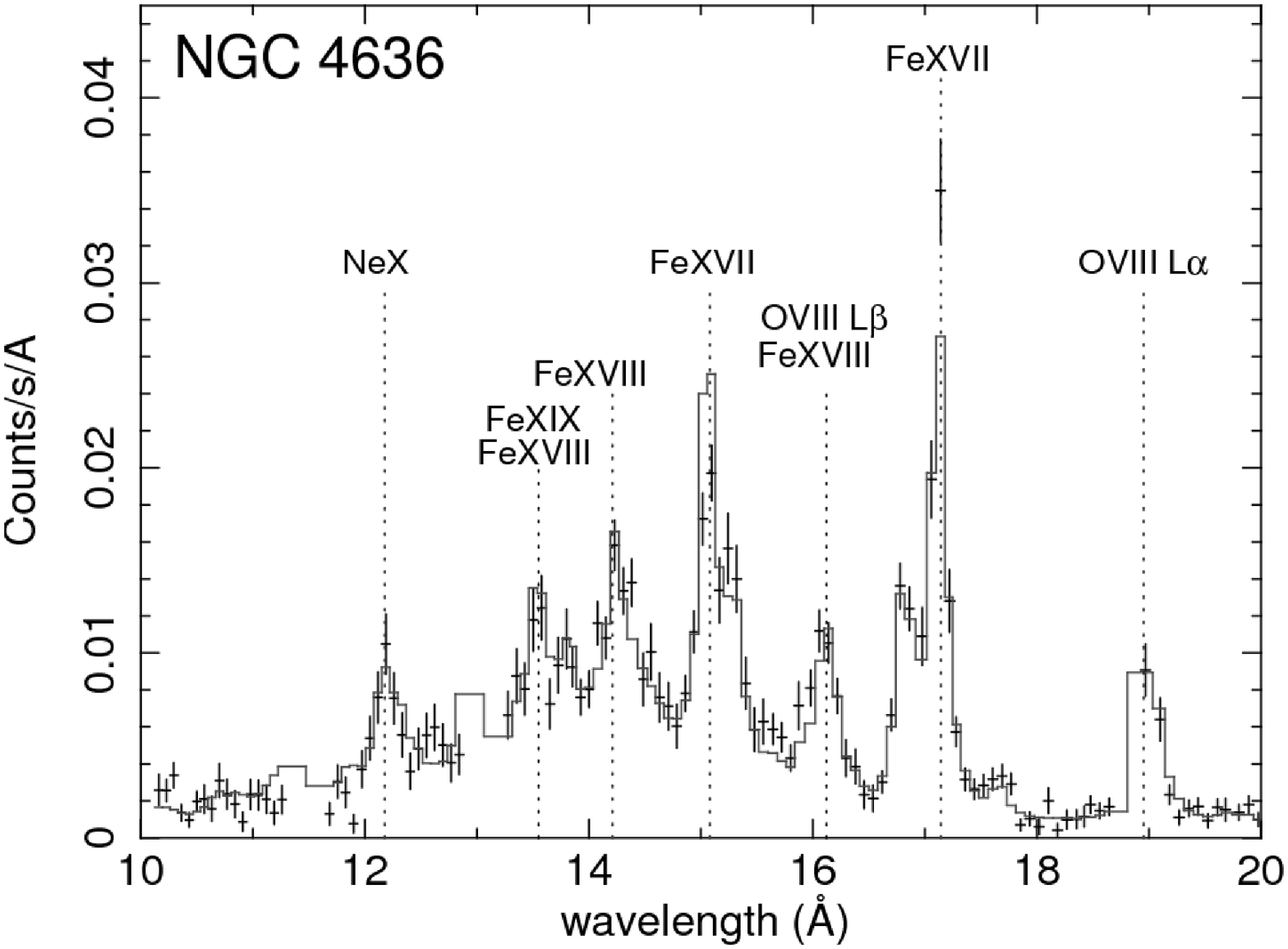}
\end{minipage}\hspace{25mm}
\begin{minipage}{0.45\textwidth}
  \includegraphics[width=0.95\textwidth]{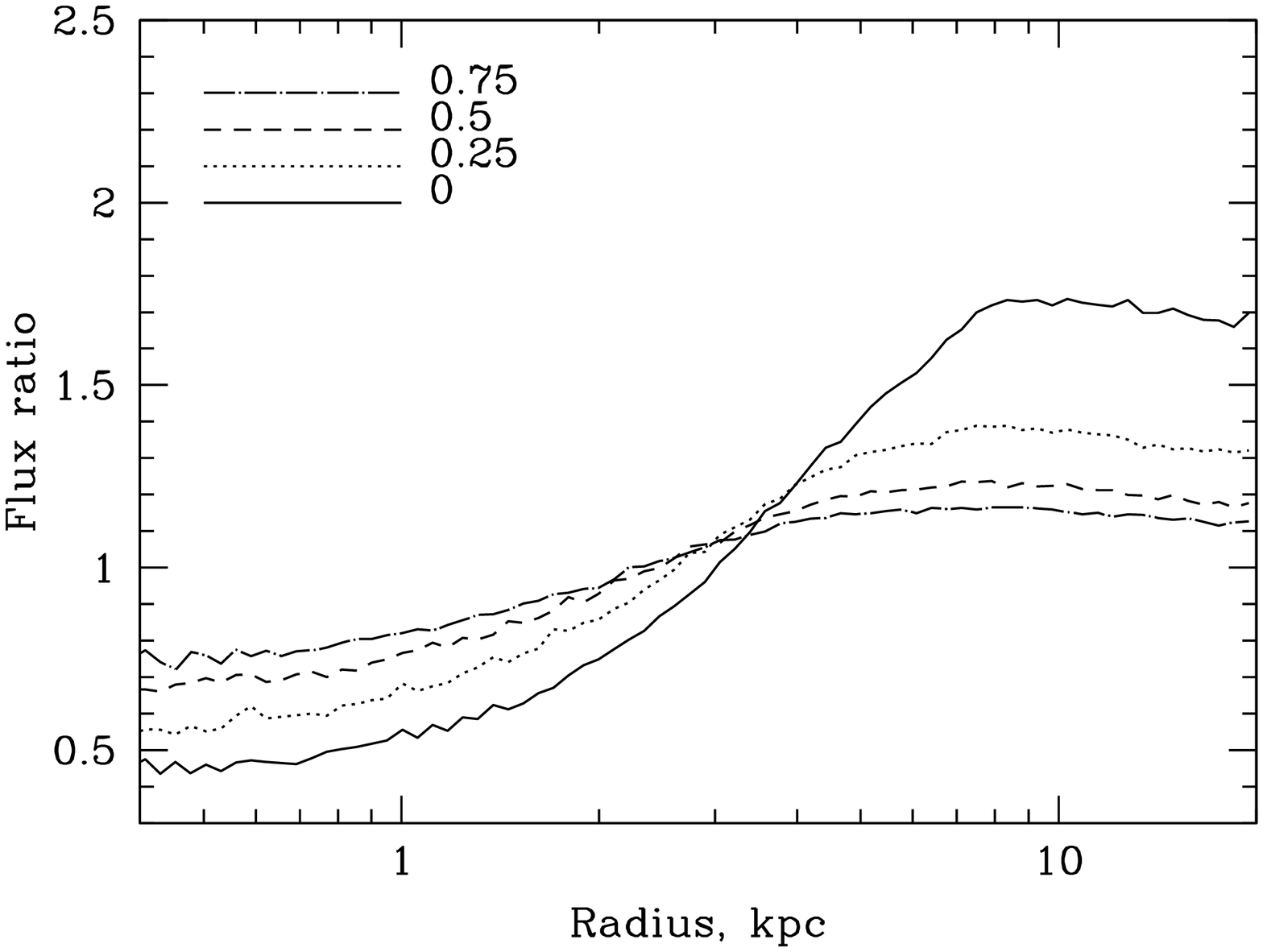} 
\end{minipage}
\vspace{-10mm}
\caption{
\footnotesize
{\it Left:} \xmm\ RGS spectra extracted from a
  0.5\arcmin\ wide region centered on the core of
  NGC~4636. {\it Right:} Simulated radial profiles of the ratio of the
  15.01~\AA\ line intensities calculated with and without the effects
  of resonant scattering, for isotropic turbulent velocities
  corresponding to Mach numbers 0.0, 0.25, 0.5, and 0.75 and a flat abundance
  profile. Adapted from \citep{Wer09}.}
\label{fig:wer09}       
\end{figure*}

\begin{figure*}
\begin{minipage}{0.44\textwidth}
  \includegraphics[width=0.95\textwidth]{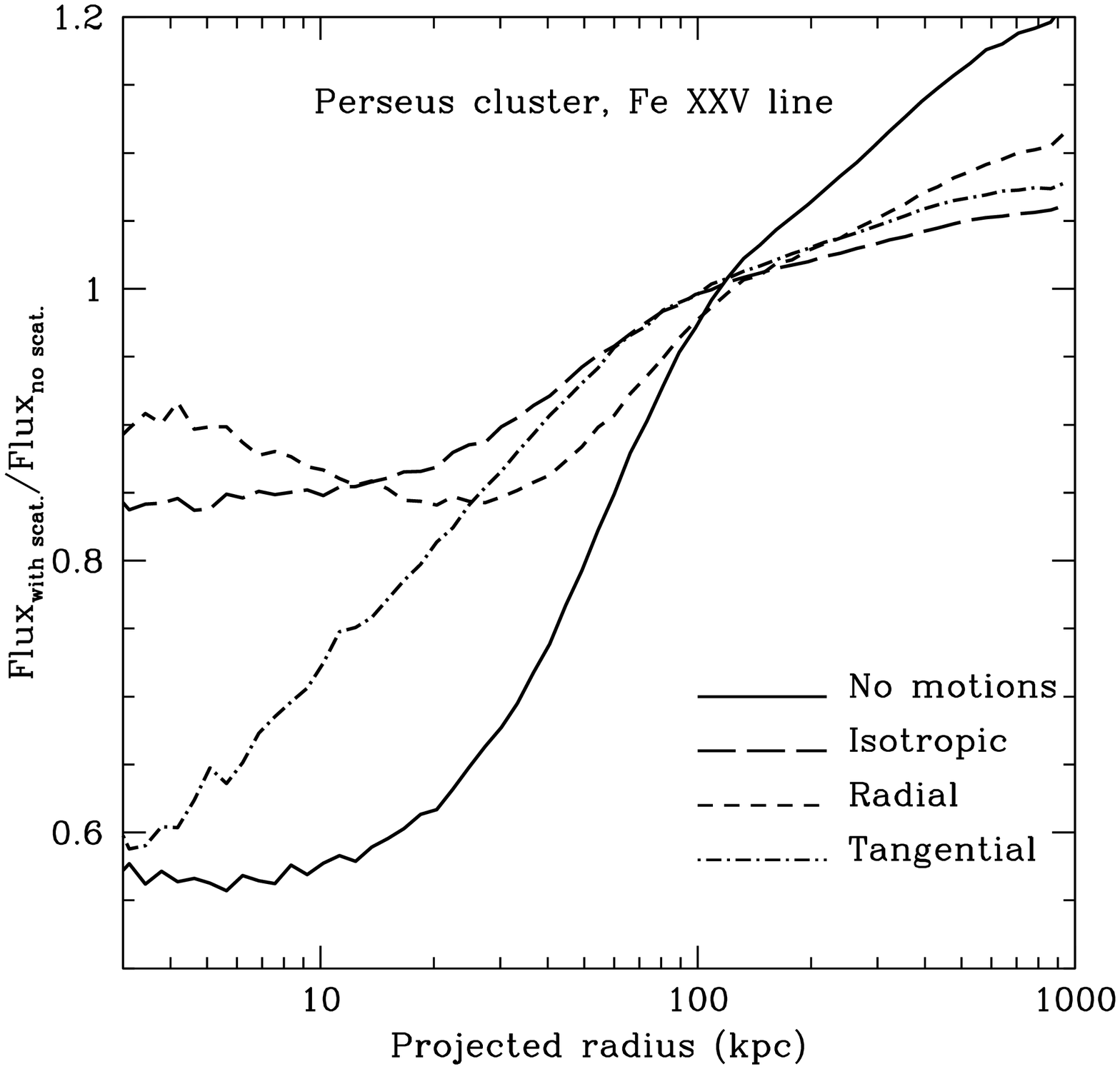}
\end{minipage}\hspace{19mm}
\begin{minipage}{0.44\textwidth}
  \includegraphics[width=0.95\textwidth]{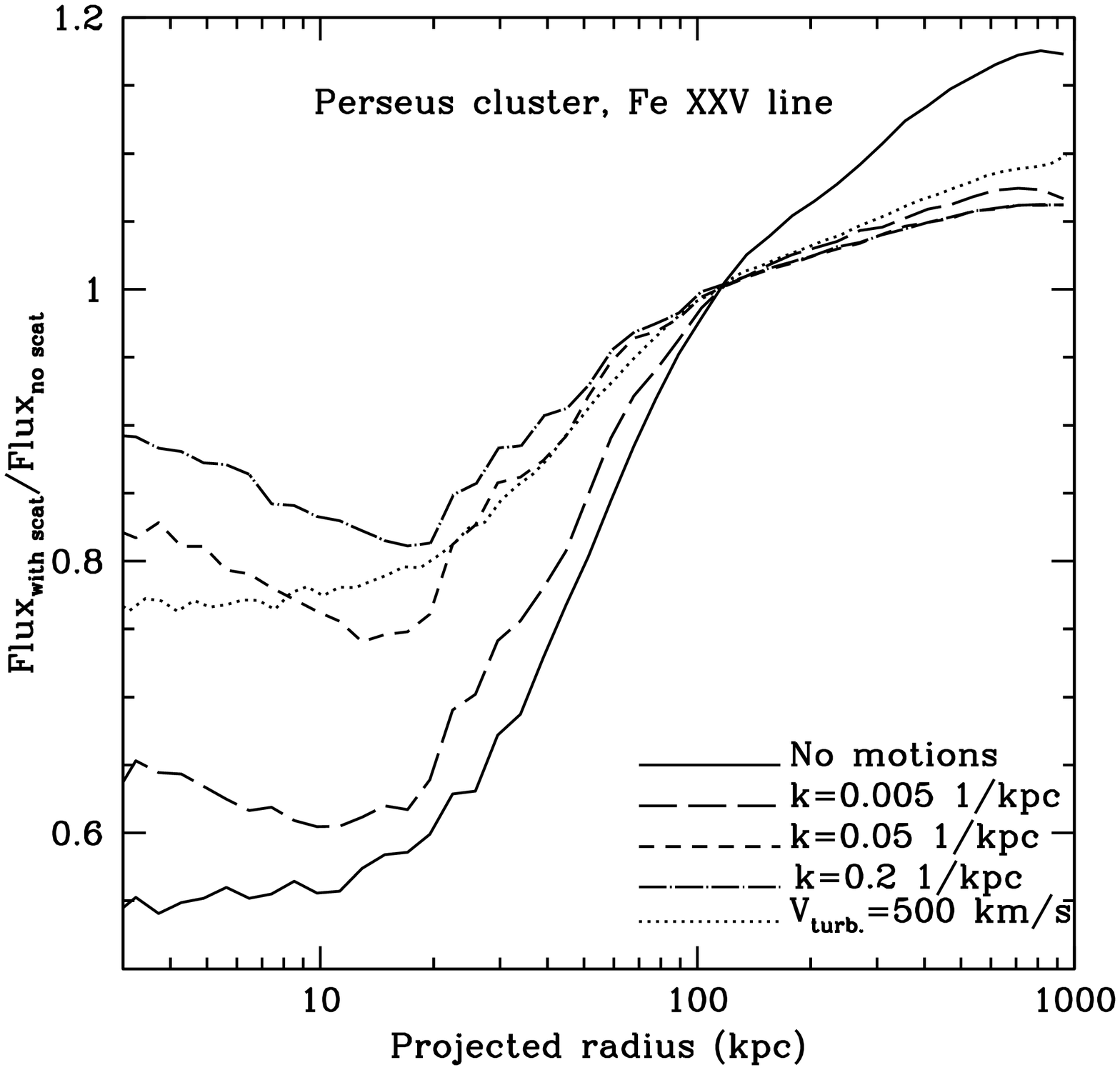} 
\end{minipage}
\vspace{-3mm}
\caption{
\footnotesize
{\it Left:} Impact of the anisotropy of stochastic gas
  motions on the ratio of the fluxes in the Fe XXV line at 6.7 keV in
  the Perseus cluster calculated with and without RS effect. {\it
    Right:} Impact of the spatial scales of gas motions on RS. Two cases of small scale motions are shown with
  dot-dashed and short-dashed curves. Case of large scale motions is
  shown with long-dashed curve. Adapted from \citet{Zhu10b}.}
\label{fig:Zhu10b}       
\end{figure*}

RS in the brightest X-ray emission lines can cause
distortions in the surface brightness distribution, changes in line
spectral shapes, variations in the abundance of heavy elements and
polarization in lines. The magnitude of these effects is 
also sensitive to the characteristics of the gas velocity field
(\citet{Gil87}, see also review \citet{Chu10}).

If one accurately measures the line ratios of (optically thick)
resonant and (optically thin) non-resonant lines, the velocity
amplitudes of gas motions can be found. \cite{Wer09} obtained high
resolution spectra of the elliptical galaxy NGC 4636 using the
RGS on the XMM-Newton satellite. The Fe XVII line at 15.01
\AA ~is suppressed (Fig. \ref{fig:wer09}) only in the dense core and 
not in the surrounding regions, while the line of Fe XVII at 17.05 \AA ~is
optically thin and is not suppressed. \citet{Wer09}
modeled the radial intensity profiles of the optically thick line, 
accounting for the effect of RS for different values of
the characteristic turbulent velocity. Comparing the model to the
data, it was found that the isotropic turbulent velocities on spatial
scales smaller than $\approx$ 1 kpc are less than 100 km/s and the
turbulent pressure support in the galaxy core is smaller than 5 per
cent of the thermal pressure at the 90 per cent confidence level
(taking into account only statistical errors).

The relation between the RS effects and the velocity
field can be used to test anisotropy and spatial scales of gas motions. \cite{Zhu10b} showed that
RS is the most sensitive to the radial small scale
motions (Fig. \ref{fig:Zhu10b}).

\begin{figure}
  \includegraphics[width=0.39\textwidth]{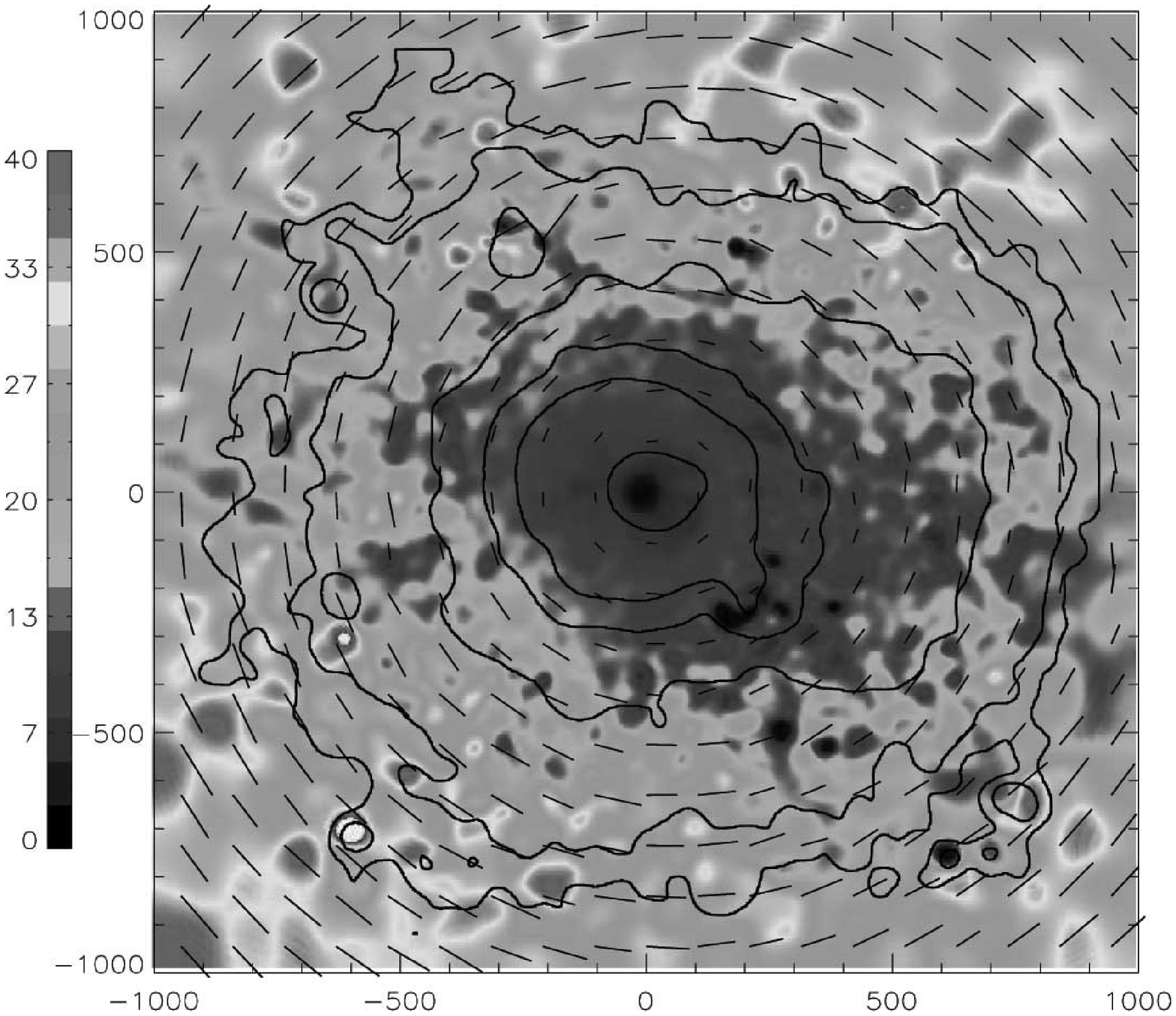}
  \includegraphics[width=0.39\textwidth]{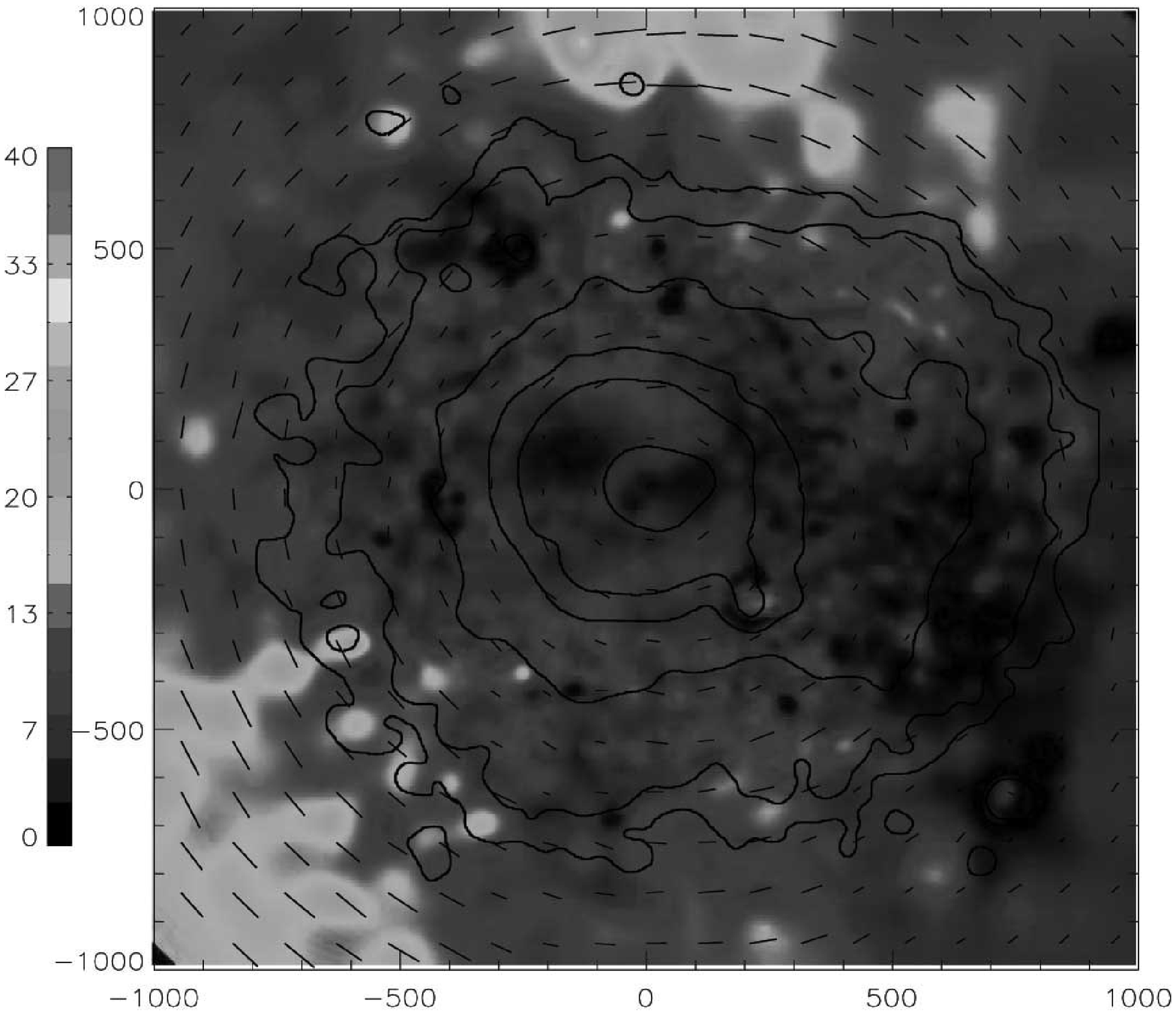}
\vspace{-2mm}
\caption{
\footnotesize
Polarization degree of the simulated massive cluster g8
  \citep{Dol08} in the line of Fe XXV at 6.7 keV.  The
  colors denote the polarization degree in per cent. The
  short dashed lines show the orientation of the electric
  vector. Contours of the X-ray
  surface brightness in the chosen line are superposed. The size of each
  picture is 2$\times$2 Mpc. 
The top panel shows the case of gas
  being at rest. The bottom panel shows the case of the gas velocities
  obtained in the simulations.  
Adapted from \citet{Zhu10a}
} 
\label{fig:3Dpol}       
\end{figure}

RS also gives us a unique opportunity to put
constraints on transverse gas motions in galaxy clusters by means of
polarization in lines (see  \cite{Saz02, Zhu10a}). 
Calculated polarization degree is shown in
Fig. \ref{fig:3Dpol} \citep{Zhu10a} with and without account for gas motions.
 For one of the most massive simulated clusters (g8 in the sample of
\citet{Dol08}) the polarization degree in the 6.7 keV He-like iron line reaches
$\sim$ 25 per cent at a distance of $\sim$ 500 kpc from the center if
no turbulent motions are present (top
panel). Inclusion of
gas motions substantially decreases the polarization down to $\sim$ 10 
per cent (bottom panel) within the same region.

\section {Conclusions and Future Prospects}

We presented a convenient way to put constraints on the amplitude and shape of
3D PDS using only observables, i.e. mean velocity and dispersion along
the line-of-sight. SPH simulations show that 3D velocity PDS (i)
is distance dependent, (ii)
deviates significantly from the canonical Kolmogorov PDS. This results should be
verified with AMR simulations and simulations with higher resolution.

Line-of-sight velocity and dispersion measurements will be available in the nearest future with {\em  Astro-H} mission. 
We discussed RS as one of the ways to study gas motions with {\em
  Astro-H} data, showing that RS is most sensitive to radial small
scale gas motions. Moreover future polarimetric mission
(e.g. on-board of {\em IXO}) will fully exploit RS
and allow us to constrain transverse gas motions. 

\begin{acknowledgements}
I am grateful to my collaborators Eugene Churazov, Rashid Sunyaev,
Klaus Dolag, Norbert Werner, Sergey Sazonov and William Forman. 
I would like to thank the International
Max Planck Research School (IMPRS) in Garching.
\end{acknowledgements}

\bibliographystyle{aa}

\begin{thebibliography}{}

\bibitem[Ar{\' e}valo et al. (2010)]{Are10} Ar{\' e}valo P.,  Churazov
  E., Zhuravleva I., Hern{\' a}ndez-Monteagudo C., Revnivtsev M., 2010, ApJ, submitted

\bibitem[Churazov et al. (2004)]{Chu04} Churazov E., Forman W., Jones C., Sunyaev 
R., B{\"o}hringer H., 2004, MNRAS, 347, 29 

\bibitem[Churazov et al. (2010)]{Chu10} Churazov E., Zhuravleva I., Sazonov S., 
Sunyaev R., 2010, SSRv, 104 

\bibitem[Dolag et al. (2008)]{Dol08}
    Dolag K., Borgani S., Murante G., Springel V., 2009, MNRAS, 399,
    497

\bibitem[Gilfanov, Syunyaev \& Churazov (1987)]{Gil87} Gilfanov M.~R., Syunyaev R.~A., Churazov E.~M., 1987, PAZh, 13, 7 

\bibitem[Inogamov \& Sunyaev (2003)]{Ino03} Inogamov N.~A., Sunyaev R.~A., 2003, AstL, 29, 791 




\bibitem[Sanders, Fabian \& Smith (2002)]{San10} Sanders J. S.,
  Fabian A. C., Smith R. K., 2010, MNRAS, in press, arXiv:1008.3500

\bibitem[Sazonov, Churazov \& Sunyaev (2002)]{Saz02} Sazonov S.~Y., Churazov E.~M., Sunyaev
  R.~A., 2002, MNRAS, 333, 191


\bibitem[Werner et al. (2009)]{Wer09} Werner N., Zhuravleva I., Churazov E., Simionescu A., Allen S.~W., Forman W., Jones C., Kaastra J.~S., 2009, MNRAS, 398, 23 

\bibitem[Zhuravleva et al.(2010)]{Zhu10a} Zhuravleva I.~V., Churazov E.~M., Sazonov 
S.~Y., Sunyaev R.~A., Forman W., Dolag K., 2010a, MNRAS, 403, 129 

\bibitem[Zhuravleva et al. (2010b)]{Zhu10b} Zhuravleva I.~V., Churazov E.~M., Sazonov 
S.~Y., Sunyaev R.~A., Dolag K., 2010b, Astronomy Letters, in press


\end{thebibliography}

\end{document}